\documentstyle[twocolumn,prc,aps,epsf,epsfig]{revtex}
 \begin{document}
 
\newcommand{\be}{\begin{eqnarray}}
\newcommand{\ee}{\end{eqnarray}}
\twocolumn[\hsize\textwidth\columnwidth\hsize\csname@twocolumnfalse\endcsname
\title{Dependence of hadronic properties on Quark
Masses and
Constraints on their Cosmological Variation
}

\author{ V.V. Flambaum$^1$ and E.V. Shuryak$^2$ }
\address{$^1$
 School of Physics, The University of New South Wales, Sydney NSW
2052,
Australia and Institute for Atomic and Molecular Theory, Harvard-Smithsonian
Center for Astrophysics, 60 Garden st, Cambridge, MA 02138-1516, USA. 
}
\address{$^2$ 
Department of Physics and Astronomy, State University of New York, 
Stony Brook NY 11794-3800, USA
}

\date{\today}
\maketitle

\begin{abstract}
  We follow our previous paper on possible cosmological
variation of weak scale (quark masses) and strong scale, inspired by 
data on cosmological variation of the electromagnetic
fine structure constant from  distant quasar (QSO)
absorption spectra. In this work we identify the
{\em strange quark mass} $m_s$ as the most important quantity, and the
{\em sigma meson mass} as the ingredient of the nuclear forces
most sensitive to it. As a result, we claim significantly
stronger limits on ratio of weak/strong scale ($W=m_s/\Lambda_{QCD}$)
 variation following from our previous discussion of
primordial Big-Bang Nucleosynthesis ($|\delta W/W|<0.006$) and
 Oklo natural nuclear reactor ($|\delta W/W|<1.2 \cdot 10^{-10}$; 
there is also a non-zero solution
 $\delta W/W=(-0.56 \pm 0.05) \cdot 10^{-9}$) .
\end{abstract}
\vspace{0.1in}
]
\begin{narrowtext}
\newpage

\section{Introduction}
  The understanding of how exactly the fundamental parameters
of the Standard Model enter any observable is certainly 
one of the most important aims of hadronic/nuclear physics.
Two old deep questions drives its discussion:
(i) Can there be ``alternative Universes'' with different
set of parameters, and what are the boundaries of the World we know
in the parameter space?; (ii) How to observe cosmological variations
of weak and strong scales? 

Discussion of both issues has been significantly
revived recently. We will not
 go into questions (i)
(see
e.g. \cite{alternative_universes}) and only mention the latter (ii).
 The issue of cosmological time variation of  major constants of physics
has been recently revived by
 astronomical data which seem to suggest a variation
 of electromagnetic $\alpha$  at the $10^{-5}$ level
 for the time scale 10 bn years, see \cite{alpha}.
The statistical significance of the effect at the moment
obviously exclude any random fluctuations, so the effect
definitely exists. Whether it may or may not have a conventional
explanation is not yet clear: more experimental work is clearly
needed
to reach any conclusions.

Nevertheless, it is quite timely to have another look at
existing limits on time variation of all the fundamental constants.
In particular, since the electromagnetic and weak forces are mixed together
in the Standard Model, one may expect a similar modification of the weak
couplings, the weak scale in general and quark masses in particular.
In fact, one can measure only variation of dimensionless parameters.
Therefore, we obtain limits on variation of $m_s/\Lambda_{QCD}$ where $m_s$ is
 the strange quark mass and $\Lambda_{QCD}$ is the QCD scale
defined as a position of Landau pole in logarithm  for running coupling
constant. It is convenient to put $\Lambda_{QCD}=const$.

A generic further argument goes as follows. The masses of 3 heavy quarks
-- c,b,t -- are {\em too large} to be important in hadronic and nuclear
physics. The masses of two light ones -- u,d -- are important, in
particular
via the pion contribution to nuclear forces
 we studied in our previous paper \cite{FS} and a subsequent one
 \cite{Dmitriev}.
 Much more extensive discussion of this issue in  
 a  contexts of chiral perturbation theory can be found in
literature, see e.g. \cite{Beane_Savage} and references therein. 
The conclusion reached in those studies is that the $m_u,m_d$ are  {\em
too small} to be really important.

Thus,  we focus on the dependence on the $strange$
quark mass $m_s$, the only one which  has the right  magnitude to
generate a maximal
sensitivity of the hadronic/nuclear physics
to the weak scale. Indeed, as follows from
QCD phenomenology, its variation from 0 to experimental value
influence
vacuum parameters such as the quark condensates $\bar q q$ at the
factor-2 level. It also affects the masses of even
$non-strange$ hadrons such as the nucleon
at about 20 percent level, etc. The fundamental explanation
of why such unexpected ``strangeness'' 
 is in fact present is related with the important role
of the instanton-induced effects in
QCD. As it follows from the chiral anomaly relation and is explained in details
e.g. in a review \cite{SS_98}, the multi-fermion
't Hooft interaction necessarily involves all 3 flavors at all steps,
even in the interactions of light u,d quarks. 

 The main new element of the present work is identification of the
 specific hadron, the
$ \sigma$, now known as $f_0(600)$ meson in the Review of Particle Properties \cite{RPP}, as the 
important ingredient of the nuclear physics most sensitive 
to $m_s$. This  resonance is known for a very long time
as a structure in $\pi\pi$ scattering, 
and is widely used in nuclear physics, see  e.g. a review
on applications of Walecka model \cite{SW}. And nevertheless,
its acceptance as a real hadronic
 state was difficult and
had a very complicated history. Non-relativistic
quark models of hadrons intensify it is as a l=1 or p-state, 
and tend to predict its mass to be around 1.4 GeV or so, in striking
contradiction to
 experiment which finds a mass of only about $m_\sigma \sim
 500\, MeV$. Also large width 
of $\sigma$ makes this state to be easily deformed by all kind of
 effects,
see e.g. recent work by one of us predicting drastic modification
on the sigma shape in pp and heavy ion collisions \cite{SB}.
On the other hand, recent data have elucidated sigma production in
a set of much simpler situations, from heavy quark hadrons, $\Upsilon$
 transitions
and $D$ decays. Those consistently point toward the smaller width
 $\Gamma_\sigma \sim 250 \, MeV$. This reduced the controversy and
 put this particle back into the
Review of Particle Properties \cite{RPP}. 

In the next section we will argue that the sigma mass
has  stronger sensitivity to $m_s$, than that of ordinary non-strange
hadrons
like $N$ or $\omega$. This happens because of
 a valence strange part plus the
repulsion from the nearby $\bar K K,\eta\eta$ continuum thresholds.
We then estimate the derivative of the deuteron binding and neutron
resonance energies to $m_s$.

\section{Sensitivity of hadronic masses to variation of quark masses}
\subsection{Why sigma?}
 The very first appearance of the sigma mesons was as a two-pion 
scalar-isoscalar resonance. It has been gradually learned that
the corresponding channel for $\bar q q$ 
interaction is the ``maximally attractive channel'', with attraction
so strong that it breaks spontaneously chiral symmetry and produce
the non-zero quark condensate. The mechanism 
of that attraction is attributed mostly to instanton-induced
't Hooft interaction, for a review see \cite{SS_98}.

  Sigma meson is an excitation on top of the scalar condensate, a kind of a
  Higgs boson of strong interactions. If indeed one naively assumes that
it underlines all hadronic masses, e.g. that of the nucleon, the
corresponding coupling can be estimated as
\be \label{eqn_gs_naive}
g_s=M_n/f_\pi\approx 10 \ee. This large value in turn implies that
the perturbation theory can only be used as a qualitative guide, at best.

In passing, let us mention that arguments about development of the
most optimum effective description of hadronic/nuclear physics
in terms of mesonic degrees of freedom, known also as Quantum Hadro
Dynamics,
are still going on. Using some field variables can be better than others: 
in this respect let us mention the paper 
\cite{CEG} which emphasized the instead of the traditional
$\sigma$
of the linear sigma-model, a chiral partner of the pion, one better
use the radial field $\sqrt{\sigma^2+\vec \pi^2}$ which has normal
derivative coupling to pions. There is extensive literature on loop
corrections and related observables, such as resonance mass and shape
modification in nuclear and excited hadronic matter, see
e.g. \cite{SB}
as a recent example.

For the purposes of this work
it would be sufficient to use
  simple and  widely used Walecka model,
which keeps
only the sigma and the omega exchanges in the effective nuclear forces
\be \label{walecka}
V= -{g_s^2 \over 4\pi} {e^{-r m_\sigma} \over r}+{g_v^2\over 4\pi}
 {e^{-r m_\omega} \over r}\ee
The very important lesson about nuclear forces this model had
emphasized is that the
nuclear potential  is in fact a highly tuned small difference
 of
two large terms. 

We will argue in the next section that there are reasons to think that
the sensitivity of these two terms to the fundamental weak scale
is quite different. Sigma (scalars) involve all quark flavors
while omega (vectors) do not, forbidden by Zweig rule.
As a result, {\em there is no  fine tuning in the derivative 
   over $m_s$}, which 
significantly enhances the effect to be derived.

The values of the two coupling constants used in the nuclear
 matter
applications of this model \cite{SW} are
\be\label{walecka_gs}  g_s^2= 357.4 m_\sigma^2/m_N^2  \approx 100
 \nonumber \\
   g_v^2=273.8 m_\omega^2/m_N^2 \approx 190 \ee 
for $m_\sigma=500 \, MeV$. Note that the effective scalar coupling
is close to naive value (\ref{eqn_gs_naive}) mentioned above.

  \subsection{ Strange valence and strange sea  of the $\sigma$ meson }

Scalar and pseudoscalar mesons are different from more familiar
vector and axial ones in terms of their flavor composition.
 In the latter case the so called Zweig rule applies, forbidding
flavor mixing: so for example the $\omega$ meson we will discuss in his
work has a truly negligible mixing with a strange counterpart, $\phi$ meson.
Scalar and pseudoscalar mesons are on the contrary nearly ideally
$SU(3)$ octets and singlets, so
 different flavors are very  strongly mixed together.
The pseudoscalar channel has been studied  extensively,
and we know that in this case $\eta'$ is 
 much heavier than (twice more strange) $\eta$ meson.
This is the famous Weinberg U(1) problem, resolved by
existence of the instanton-induced
repulsive interaction pushing the singlet state upward.

 The same instanton-induced interaction has the {\em opposite sign} in the $SU(3)$ singlet
scalar channel, pushing $m_\sigma$ downward, see \cite{SS_98}
for details. The magnitude of flavor mixing matrix element
 in the scalar channel
has been also evaluated on the lattice,  and  the
results
agree with the instanton-based predictions in sign and magnitude.

Apart of theoretical motivation, there is a simple phenomenological fact
that no purely  strange $\bar s s$ counterpart to sigma resonance $f_0(600)$ 
is seen. There are strong evidence that the
 pair of states $f_0(980),a_0(980)$ are the $\bar K K $ molecule, 
so the next  $f_0$ are at 1300 and 1500 MeV.

For those reasons, we  think that
 a description of $\sigma$ as a 
$SU(3)$ singlet  state
\be \sigma = \frac{1}{\sqrt{3}}({\bar u u + \bar d d +\bar s s}) \ee
strongly split from the $SU(3)$ octet one
is a reasonable  approximation.
This means that the valence contribution to the derivative is
\be \label{valence} 
{\partial  m_\sigma\over \partial m_s}|_{val}= <\sigma|\bar s
 s|\sigma>=2/3\ee
since with probability 1/3 there is a strange $pair$.
 A mixing between different scalar
mesons $f_0$ ($\sigma\equiv f_0(600)$,  $f_0(980)$, $f_0(1370)$)
would further change  the valence contributions downward. However,
based on large gap between sigma and other states, and also based on
better studied $\eta\eta'$ mixing, one might think
that the relative change due to mixing
with next $f_0$ states is not significant.

 Let us now consider the contribution of the so called strange sea,
virtual $\bar s s$ pairs, which are always present even in
a completely non-strange hadrons like a nucleon.
The relatively well studied case is
the nucleon mass sensitivity to $m_s$, 
 already  discussed in \cite{FS}. Let briefly remind
that  $KN$ scattering data imply that
\be \label{NssN}
{\partial m_N \over \partial m_s}=<N|\bar s s |N>\approx 1.5 \ee
and thus
 about 1/5 of the nucleon
 mass comes from the strange
 sea ($m_s \approx$ 120 MeV). 

Similar matrix elements for $\sigma,\omega$ mesons is not
possible to obtain experimentally, although it can be done
on the lattice. To estimate it we will adopt a simple
constituent quark picture, assuming
 {\em additivity} of the strange sea. If so, the 
 derivatives analogous to (\ref{NssN}) for all mesons
 should be $2/3$ of it, or
\be \label{meson_sea}
{\partial m_{mesons} \over \partial m_s}|_{sea}\approx 1 \ee
As we will see later, the exact value of the common sea contribution
is not actually important, since its contributions to $\sigma$
and $\omega$ mesons tend to cancel each other nearly exactly
when we calculate variation of $N-N$ interaction.
What matters is the $difference$  between their strange seas, to be
discussed in the next section.

It is convenient
to present the effect of the possible quark mass variation on the
 $\sigma$ mass in the following form
\be \frac{\delta m_\sigma}{ m_\sigma}= 0.4 (\frac{\delta
  m_s}{m_s} + \frac{m_u +m_d}{m_s}\frac{\delta m_q}{ m_q})=
 0.4\frac{\delta  m_s}{m_s} + 0.04 \frac{\delta m_q}{ m_q}
 \ee
where we have used $m_s =120\, MeV$.
We see that the relative change of the strange quark mass
produces much larger effect than the relative change of the 
light quark mass. This is similar to the case of the nucleon mass
variation
\be \frac{\delta m_N}{ m_N}= 0.19 \frac{\delta
  m_s}{m_s} + 0.045 \frac{\delta m_q}{ m_q} \ee

\subsection{$\bar K K,\eta\eta$ mixing with $\sigma$ }

  As the second approximation, we will discuss loop effects, 
or mixing with 2-meson states. Those are also completely
different for $\sigma$ and $\omega$ mesons.
As we mentioned earlier, the latter practically does not mix with
 $\bar K K$ states, while  $\sigma$ does mix with them strongly.
An admixture of virtual $\bar K K,\eta\eta $ pairs can be viewed
as an additional contribution to the strange sea, on top of the
strange content of the non-strange constituent quarks (\ref{meson_sea}).

The $\sigma$  mixing with {continuum} of pseudoscalars is described by 
the standard
 mass operator  given by the usual loop diagram 
\be \Sigma(Q^2)= \int {d^4k \over 2\pi^4} {\lambda^2(k,Q) \over
  [(k+Q/2)^2-m^2+i\epsilon]  [(k-Q/2)^2-m^2+i\epsilon]}\ee
where $\lambda$ is the $KK\sigma,\eta\eta\sigma$ or $\pi\pi\sigma$ couplings.
Its real part describes the shift of $m_\sigma$ due to
repulsion from these states: the imaginary part
(for pions only) gives the width. The sign 
of the shift due to 
$\bar K K$ is obviously negative, since $m_\sigma<2m_K$.
The effect of $\pi\pi$ has contributions of both signs.
For constant coupling the 
shift
is logarithmically divergent, in reality
it has to be regulated by formfactors in the
vertexes. The total shift is negative and large, of the order
of very large width $\Gamma_\sigma\sim 300 \, MeV$. 
Note that this large negative shift is partly the reason why
 the  $\sigma$ mass is so small.
However, in this paper we focus on the 
dependence on quark masses. Assuming that the main dependence
comes from masses of Goldstone bosons, $m_\pi,m_K$, we differentiate
the mass operator over these masses and obtain the 
convergent result. Thus one can ignore formfactors and
extract the effective coupling constant out of the integral.

For the derivative at $Q=m_\sigma=0.5 \, GeV$
 we get the following numerical value for the shifts
\be {\partial \Sigma \over \partial m_K^2}= 0.0229 GeV^{-2}\lambda^2_{\sigma KK}
\ee
\be {\partial \Sigma \over \partial m_\eta^2}= 0.019 GeV^{-2}\lambda^2_{\sigma \eta\eta}
\ee
The couplings are not experimentally known, so we rely on the SU(3) symmetry
and relate them to $\lambda_{\sigma \pi\pi}$, which is in turn related to
sigma meson width
\be \Gamma_\sigma = {3\over 2} { \lambda_{\sigma \pi\pi}^2 \over 16\pi
 m_\sigma} \sqrt{1-{4 m_\pi^2 \over m_\sigma^2}} \ee
Taking $\Gamma_\sigma$=250 MeV we obtain  $\lambda_{\sigma \pi\pi}^2$=
 5 Gev$^2$.
The factor 3/2 account for  $\pi^+\pi^-,\pi^0\pi^0$ modes. However in the 
mass shift there are contribution of $K^+K^-,\bar K^0  K^0,\eta\eta $
channels which we would count as 5/2. 
Substituting numbers and using standard Gell-Mann-Oaks-Renner
expressions
for  $m_\eta,m_K$ ($m^2 \propto m_s \Lambda_{QCD}$),
we obtain the 
additional sensitivity of the sigma mass shift arising
from  the mixing effects 
\be \label{mixing}
{\delta  m_\sigma\over m_\sigma}=  {  \delta m_s \over m_s}\left[2 {m_K^2
 \over 2 m_\sigma^2} {\partial \Sigma \over \partial m_K^2}+ 0.5 {m_\eta^2
 \over 2 m_\sigma^2} {\partial \Sigma \over \partial m_\eta^2}\right]
\approx   {\delta m_s \over m_s} 0.14 
 \ee
We included $K^+K^-,\bar K^0 K^0$ modes with the coefficient 1 and
 $\eta\eta$
with 1/2: the latter only contribute about 1/5 of the final answer.
We ignored even smaller contribution of the $\eta'\eta'$ loop.

\subsection{The total sensitivity of $m_\sigma$ as compared to $m_N$}

Together with the one estimated in the previous section, it leads to
total
\be {\delta  m_\sigma\over m_\sigma}\approx (0.24+0.16+0.14)
{\delta m_s \over m_s}
=0.54  {\delta m_s \over m_s}
 \ee
where 3 terms are the contributions
of the common strange sea (\ref{meson_sea}), valence strangeness (\ref{valence}) and
the loop mixing (\ref{mixing}), respectively.

In the same units the sensitivity of the nucleon mass is
\be \label{MnMs} {\delta m_N \over m_N } \approx 0.19  {\delta m_s \over m_s}
 \ee
We conclude that the sigma mass is about 3 times more sensitive to
the variation of the strange quark mass than the nucleon mass.

   We also need the sensitivity of $\omega$ meson to the strange mass
variation. This meson does not have valence strange quarks and practically
does not mix with $\phi$, $K$ and $\eta$ mesons. Therefore, only
 the strange see contributes:

\be {\delta  m_\omega\over m_\omega}\approx 0.15  {\delta m_s \over m_s}
 \ee

\section{The modification of the deuteron  binding}
\subsection{Preliminary analytic estimates}

Simple analytic estimate for sensitivity of the deuteron binding
to sigma and omega mass modification is obtained by the
differentiation of the potential over the mass and   averaging
the resulting expression $\sim exp(-mr)$  over the 
 radial wave function.

The simplest short range approximation leads to the
 free motion wave function
\be \label{naive_wf}
\psi(r)= {\sqrt{2\kappa} \over r}exp(-\kappa r)\ee
This wave function tends to infinity at small distances.
However, the real deuteron wave function should be small there
because of the repulsive core in the potential $V(r)$.
Therefore, we introduced a small cut-off radius  $b$ in the integration
over $r$. We estimated $b$  from the condition
$V(b)=0$ which gives $b=0.45 $ fm. We get
 the following shift of the deuteron binding 
\be {\delta Q_d\over Q_d}=-{ m_\sigma \over  Q_d} {\delta m_\sigma\over  m_\sigma} {g_s^2 \over 4\pi} {2\kappa \over
  2\kappa +m_\sigma} e^{-(2\kappa +m_\sigma)b} \approx
-75 {\delta m_\sigma\over  m_\sigma}\ee
 A variation of $m_{\omega}$ gives
\be {\delta Q_d\over Q_d}={ m_\omega \over  Q_d} {\delta m_\omega\over 
 m_\omega} {g_v^2 \over 4\pi} {2\kappa \over
  2\kappa +m_\omega}  e^{-(2\kappa +m_\omega)b} \approx
 80 {\delta m_\omega\over  m_\omega}\ee

The sign difference between these two derivatives and
very large derivative value
 is due to fine tuning between omega
and sigma terms, which are separately much larger  than the sum.

 The next step one can do analytically is to add
 a simplest square potential to the hard core.
  The energy of
a shallow level  in such potential is equal to \cite{landau}:
\begin{equation}\label{22}
E_d=-Q_d=-\frac{\pi^2}{16}\frac{(U-U_0)^2}{U_0},
\end{equation}
\begin{equation}\label{23}
U_0=\frac{\pi^2}{8ma^2}.
\end{equation}
Here $U$ and $a$ are the depth and width of the potential well ($a=c-b$,
where $c$ and $b$ are outer and inner radii),
$m=m_N/2$ is the reduced mass. 
Selecting the width and depth of the well to be $a=1.6 \, fm, U_0= 40.2
\,MeV, U= 52.6
\,MeV,$ we got 
  \be {\delta Q_d\over Q_d}\approx
-81.6 e^{-m_\sigma b} {\delta m_\sigma\over  m_\sigma}\ee
\be {\delta Q_d\over Q_d}\approx 87.4  e^{-m_\omega b}{\delta
  m_\omega\over  m_\omega}\ee
By changing the core radius from $b=0$ to $b=0.4$ fm one can vary the answer
 by about factor 3. Simple exponential dependence on the core radius
 $b$
appears because of translational invariance of 1d Scredinger equation
for $r\psi(r)$.

Another effect one should consider is the modification
of the nucleon mass: its contribution  to modification of the
 deuteron binding  is 
\be \delta Q_d={\delta M_N \over M_N}  <d| {p^2 \over 2  M_N}|d >
\ee
which leads to
\begin{equation}\label{24}
\frac{\delta Q}{Q}=\frac{U+U_0}{U-U_0}\,\,\frac{\delta m_N}{m_N} \simeq
 7.7\frac{\delta m_N}{m_N} .
\end{equation}
 Although the sensitivity to the nucleon mass is much
 weaker than that for mesons, it is still
quite strong: we attribute it to the fact that the small deuteron
 binding energy is in turn
 the delicate balance between larger kinetic
and the potential energies.

\subsection{Using the Walecka potential}

  The estimates of the preceding section are given for orientation only,
and in fact one of course have to
 solve numerically the radial Schreodinger equation and obtain the
correct wave function. The one can either average the potential
derivative
over it or simply vary
 all masses involved
explicitly. We did the latter and determined the sensitivity of  $Q_d$ to
the sigma, omega and nucleon mass. 

Strictly speaking, at this point it is no longer possible to limit ourselves to
Walecka model,  with the coupling constants
(\ref{walecka_gs}), since it does not describe correctly the deuteron
binding.
In fact, by ignoring all spin-dependent forces one cannot even
separate the
spin-singlet and the spin-triplet states. The tensor forces,
attributed to pion and rho exchanges, are needed for this task.  
Instead of doing so, we have chosen to modify a bit the strength
of the omega term, reducing $g_v^2$ by a factor 0.953 as compared to
(\ref{walecka_gs}) and obtaining the correct deuteron binding. 

Our results are\footnote{In the derivative over the omega 
mass we have divided back by
  the factor 0.953, restoring the original strength
of the vector term.}

\be 
{\delta Q_d\over Q_d}\approx -48 {\delta m_\sigma \over  m_\sigma}
\approx -26 {\delta m_s \over  m_s}
\ee
\be 
{\delta Q_d\over Q_d}\approx 50 {\delta m_\omega \over  m_\omega}
\approx 7.5 {\delta m_s \over  m_s}
\ee
\be 
{\delta Q_d\over Q_d}\approx 6 {\delta m_N \over  m_N}
\approx 1.1 {\delta m_s \over  m_s}
\ee
One can see that the first two derivatives are more sensitive to exact
shape of the wave function: they agree qualitatively but not quantitatively
with the analytic estimates above. We will not show here such details
as exact and approximate wave functions, but just comment that
the difference between those explain the difference in the integrals.

  Summing all the contributions we find
\be
{\delta Q_d\over Q_d}\approx -17 {\delta m_s \over  m_s}
\ee

Using limits on Big Bang Nucleosynthesis from \cite{FS} 
\be |{\delta Q_d \over Q_d}|\, < 0.1 \ee
one  gets the final limit on the $m_s$ variation to be
\be \label{limit_BBN}
\left|{\delta (m_s/\Lambda_{QCD}) \over (m_s/\Lambda_{QCD})} \right|< 0.006
\ee

\section{ OKLO}

In this section we extract limits on $\delta m_s$ following from data on natural nuclear
reactor in Oklo active about 2 bn years ago.
 The most sensitive phenomenon (used previously for limits on 
the variations of the electromagnetic $\alpha$) is disappearance of certain
isotopes (especially $Sm^{149}$) possessing a neutron resonance close to zero
 \cite{Oklo}. Today the lowest
 resonance energy is only $E_0=0.0973 \pm 0.0002 \, eV$
is large compared to its width, so the
neutron capture  cross section $\sigma \sim 1/E_0^2$.
The data constrain the ratio of this cross section to a non-resonance
one (which was used to measure  number of neutrons emitted by the reactor).
 It therefore implies
\footnote{Of course, under assumption that the $same$ resonance 
was the lowest one at the time of Oklo reactor.}
 that these data constrain the variation of the following
ratio $\delta( E_0/E_1) $ where $E_1\sim 1 \, MeV$ is a typical
single-particle  energy scale, which may be viewed as the energy of some 1-body
``doorway'' state.

A generic expression for the level energy in terms of fundamental parameters
of QCD can be written as follows
\be 
\delta E_0=A*\delta\Lambda_{QCD}+B_q*\delta m_q + B_s*\delta m_s+
C\delta\alpha* \Lambda_{QCD} \ee
where $A,B,C$ are some coefficients. The first term is the basic QCD
term, while others
 are corrections due to modification of the quark masses and 
the electromagnetic $\alpha$.

In this section we  provide new  estimate of the $B_s$.
More specifically, we estimate the variation of the resonance energy 
 resulting from
 a modification of the sigma mass. 
The energy of the resonance $E_0=E_{excitation}-S_n$ consists of 
excitation energy of a compound nucleus, minus the neutron
 separation energy $S_n$.
This, in turn, is 
a depth of the potential well $V$ minus the
neutron Fermi energy $\epsilon_F$,
$ S_n=V-\epsilon_F$. The latter scales like $1/R^2$ if the radius of the
well is changed. The kinetic part of the
excitation energy $E_{excitation}$ scales in the same way. So,
\be E_0=E_{excitation}-S_n=E_{excitation}+\epsilon_F-V=K{\hbar^2 \over M
  R}-V\ee
where K is a numerical constant which can be found from the present
time condition $E_0\approx 0$.
The  shift of the resonance then is
\be \label{E0} \delta E_{0}=-K {\hbar^2 \over M  R}\left( 
{ \delta M \over M}+
{2 \delta R \over R} \right) -\delta V   =-V \left({ \delta M \over M}+
{2 \delta R \over R} +{\delta V \over V} \right)
\ee
Using eq. (\ref{walecka}) we can find the depth of the potential well
\footnote{Note that the suppression of N-N wave function at small separation
due to the repulsive core reduces the depth of the effective potential $V$.
 However, this effect is not so important in the ratio ${\delta V \over V}$.}
\be V= \frac{3}{4 \pi r_0^3} \left(\frac{g_s^2}{m_\sigma ^2} -
\frac{g_v^2}{m_\omega ^2} \right)
\ee
Here $r_0$= 1.2 fm is an inter-nucleon distance. Numerical estimates
shows that the contribution of the variation of $r_0$ (and the variation
of $R=A^{1/3} r_0$ in eq. (\ref{E0})) is not 
as important as the direct contribution of the $m_\sigma$ variation
 in the equation above.
This gives us
\be  {\delta V \over V}=- 8.6 \frac{\delta m_\sigma}{m_\sigma}+ 
 6.6 \frac{\delta m_\omega}{m_\omega}=-3.5 \frac{\delta m_s}{m_s}
\ee
\be \delta E_0= 1.7 \times 10^8 eV \times \frac{\delta m_s}{m_s}
\ee
We used $V$= 50 MeV in eq. (\ref{E0}).
 Comparison of this result with the the observational
limits claimed in \cite{Oklo} $|\delta E_0| < 0.02$ eV 
gives a very strong limit 
\be  |\frac{\delta (m_s/\Lambda_{QCD})}{(m_s\Lambda_{QCD})}|
 < 1.2 \times 10^{-10}
\ee
at $time\approx 1.8$ billion years ago.

 Note that the authors of the last work in \cite{Oklo}
found also the non-zero solution $ \delta E_0 = -0.097 \pm 0.008$ eV.
This solution corresponds to the same resonance moved below
thermal neutron energy. In this case
\be  \frac{\delta (m_s/\Lambda_{QCD})}{(m_s\Lambda_{QCD})}
 =-(0.56 \pm 0.05) \times 10^{-9}
\ee
The error here does not include the theoretical uncertainty.

The production of nuclei with $A>5$ during BBN is strongly suppressed
because of the absence of stable nuclei with $A=5$.
$^5$He is unstable nucleus which is seen as a resonance in n-$\alpha$
elastic scattering. The ground state lies at
0.89 MeV above neutron threshold. The variation of the strange quark mass
  may influence the position of the resonance making, for example,
 $^5$He stable. Stable $^5$He at the time of BBN would change strongly the
 primordial abundances of light elements. The estimate similar to that
we made for Sm nucleus gives us a limit 
\be  \frac{\delta (m_s/\Lambda_{QCD})}{(m_s\Lambda_{QCD})}
> - 0.006
\ee
This limit corresponds to $ \delta E >$ - 0.89 MeV at the time of BBN.

We obtained limits on variation of 
$m_s/\Lambda_{QCD}$ during
the interval between the Big Bang and present time and on
shorter time scale from Oklo natural nuclear reactor
which was active 1.8 billion years ago. It is also possible
to obtain limits on the intermediate time scale.
One possibility is related to  position of the resonance
in  $^{12}C$ during production of this element in stars.
This famous resonance at E=380 KeV  is needed
to produce enough carbon and create life. According
to Ref. \cite{c12} the position of this resonance
can not shift by more than 60 KeV (
one can also find in Ref. \cite{c12} the limits
on the strong interactions and other relevant references).
We have made a very rough estimate of the limit on the strange 
quark mass variation which can be obtained via $m_\sigma$-mechanism:
\be  |\frac{\delta (m_s/\Lambda_{QCD})}{(m_s\Lambda_{QCD})}|
 < 0.001 
\ee
This limit can be improved after an accurate calculation.

{\bf Acknowledgments}
  We thank J.Bjorken, G.Brown and V.Dmitriev for helpful discussions.
One of the authors (VF) is thankful to the State University of New York
and Institute for Theoretical Atomic and Molecular Physics,
 Harvard-Smithsonian Center for Astrophysics for hospitality and support.
This work is also supported by the Australian Research
Council.
ES work is partially supported by the US-DOE grant
No. DE-FG02-88ER40388.

\end{narrowtext}

\begin{thebibliography}{99}
\bibitem{alternative_universes}J.D.Bjorken,  Cosmology and the standard model.
hep-th/0210202 and references therein. 
\bibitem{alpha}
  J. K. Webb , V.V. Flambaum, C.W. Churchill, M.J. Drinkwater,
 and J.D. Barrow,
 Phys. Rev. Lett., 82, 884-887, 1999.
  J.K. Webb,
M.T. Murphy, V.V. Flambaum, V.A. Dzuba, J.D. Barrow,C.W. Churchill,
 J.X. Prochaska, and A.M. Wolfe,  Phys. Rev. Lett.
87, 091301 -1-4 (2001).
 M. T. Murphy, J. K. Webb, V. V. Flambaum, V. A. Dzuba, C. W. Churchill, J.
X. Prochaska, J. D. Barrow and A. M. Wolfe,
 Mon.Not. R. Astron. Soc. 327, 1208 (2001) ; astro-ph/0012419.
M.T. Murphy, J.K. Webb, V.V. Flambaum, C.W. Churchill,
and J.X. Prochaska.  Mon.Not. R. Astron. Soc. 327, 1223 (2001);
 astro-ph/0012420.
M.T. Murphy, J.K. Webb, V.V. Flambaum, C.W. Churchill,
 J.X. Prochaska, and A.M. Wolfe.
  Mon.Not. R. Astron. Soc. 327,1237 (2001); astro-ph/0012421.

\bibitem{FS}
V.~V.~Flambaum and E.~V.~Shuryak,
Phys.\ Rev.\ D {\bf 65}, 103503 (2002)
[arXiv:hep-ph/0201303].

\bibitem{Dmitriev} V.F. Dmitriev, V.V. Flambaum, astro-ph/0209409.
\bibitem{Beane_Savage} S.R.Beane and M.J.Savage, nucl-th/0208021.
\bibitem{SS_98}
T.~Schafer and E.~V.~Shuryak,
Rev.\ Mod.\ Phys.\  {\bf 70}, 323 (1998)
[arXiv:hep-ph/9610451].
see also a comment on scalars by the same authors, hep-lat/0005025
\bibitem{SW} B.~D.~Serot and J.~D.~Walecka,
Adv.\ Nucl.\ Phys.\  {\bf 16}, 1 (1986).
\bibitem{CEG}
G.~Chanfray, M.~Ericson and P.~A.~Guichon,
Phys.\ Rev.\ C {\bf 63}, 055202 (2001)
[arXiv:nucl-th/0012013].
\bibitem{SB}
E.~V.~Shuryak and G.~E.~Brown,
arXiv:hep-ph/0211119.
\bibitem{RPP} Review of Particle Properties, Phys.Rev.D66  (2002)
  010001

\bibitem{landau} L.D. Landau, E.M. Lifshits. Quantum Mechanics,
Theoretical Physics, Volume 3 (Nauka, Moscow,1974).

\bibitem{Oklo} A.I.Shlyakhter, Nature 264 (1976) 340; T.Damour and
F.J.Dyson, Nucl.Phys.B 480 (1996) 37. Y.Fujii, A.Iwamoto, T.Fukahori,
T. Ohnuki, M. Nakagawa, H. Hidaka, Y. Oura, P. Moller.
Nucl.Phys.B 573 (2000) 377.

\bibitem{c12} H. Oberhummer, R. Pichler, A. Csoto, nucl-th/9810057.


\end{thebibliography}
\end{document}